\documentstyle[12pt,aasms4,epsfig]{article}
\catcode`\@=11
\def\gsim{\ifmmode{\mathrel{\mathpalette\@versim>}}
    \else{$\mathrel{\mathpalette\@versim>}$}\fi}
\def\lsim{\ifmmode{\mathrel{\mathpalette\@versim<}}
    \else{$\mathrel{\mathpalette\@versim<}$}\fi}
\def\@versim#1#2{\lower 2.9truept \vbox{\baselineskip 0pt \lineskip 
    0.5truept \ialign{$\m@th#1\hfil##\hfil$\crcr#2\crcr\sim\crcr}}}
\catcode`\@=12

\def\listitem{\par \hangindent=50pt\hangafter=1
     $\ $\hbox to 20pt{\hfil $\bullet$ \hfil}}

\def\puncspace{\ifmmode\,\else{\ifcat.\C{\if.\C\else\if,\C\else\if?\C\else%
\if:\C\else\if;\C\else\if-\C\else\if)\C\else\if/\C\else\if]\C\else\if'\C%
\else\space\fi\fi\fi\fi\fi\fi\fi\fi\fi\fi}%
\else\if\empty\C\else\if\space\C\else\space\fi\fi\fi}\fi}
\def\msun{\hbox{$M_\odot$}}
\def\yr-1{\hbox{${\rm yr}^{-1}$}}
\def\SP{\let\\=\empty\futurelet\C\puncspace}

\def\h1{$h^{-1}$\SP}

\def\lsim{~\rlap{$<$}{\lower 1.0ex\hbox{$\sim$}}}
\def\gsim{~\rlap{$>$}{\lower 1.0ex\hbox{$\sim$}}}
\def\void#1{{}}
\def\h{h^{-1}Mpc}
\def\pap1{paper~{\sc i}}

\def\mg2{Mg$_2$}   
\def\so{$\sigma_\circ$}
\begin{document}
\title{CLUSTER VS. FIELD ELLIPTICAL GALAXIES AND CLUES ON THEIR FORMATION
\footnote{Based on a database from observations made at the 
European Southern Observatory (ESO), La Silla, Chile;
Complejo Astronomico El Leoncito (CASLEO), San Juan, Argentina; and 
Michigan-Dartmouth-M.I.T. Observatory (MDM), Kitt Peak, U.S.A.}}

\author{Mariangela Bernardi \altaffilmark{2,}\altaffilmark{3}, 
Alvio Renzini \altaffilmark{2}, 
Luiz N. da Costa \altaffilmark{2,}\altaffilmark{4},
Gary Wegner \altaffilmark{5}, 
M. Victoria Alonso \altaffilmark{6},
Paulo S. Pellegrini \altaffilmark{4}, 
Charles Rit\'e \altaffilmark{4}, 
Christopher, N.A. Willmer \altaffilmark{4,}\altaffilmark{7}}
\altaffiltext{2}{European Southern Observatory, Karl-Schwarzschild-Str.2,
      D-85748 Garching bei M\"unchen, Germany}
\altaffiltext{3}{Universit\"{a}ts-Sternwarte M\"{u}nchen,
     Scheinerstr. 1, D-81679 M\"{u}nchen, Germany}
\altaffiltext{4}{Departamento de Astronomia CNPq/Observat\'orio
     Nacional, rua General Jos\'{e} Cristino 77, Rio de Janeiro, R.J. 20921
     Brazil}
\altaffiltext{5}{Departement of Physics and Astronomy, 6127 Wilder Laboratory,
     Dartmouth College, Hanover, NH 03755-3528, U.S.A.}  
\altaffiltext{6}{Observatorio Astron\'omico de C\'ordoba
     Laprida 854, C\'ordoba (5000), Argentina}
\altaffiltext{7}{Present address UCO/Lick Observatory, University
     of California, 1156 High St. Santa Cruz, CA, 95064, U.S.A.}

\begin{abstract}  Using new observations for a sample of 931 early-type
galaxies we investigate whether the \mg2--\so
\ relation shows any dependence on the local environment. The galaxies
have been assigned to three different environments depending on the
local overdensity: clusters, groups, and field, having used our
complete
redshift database to guide the assignment of galaxies.  It is found that
cluster, group and field early-type galaxies follow almost identical \mg2--\so
\ relations, with the largest \mg2 \ zero-point difference (clusters minus
field) being only $0.007\pm 0.002$ mag.  No correlation of the residuals
is found with the morphological type or the bulge to disk ratio.  
Using stellar population models in a differential fashion,
this small zero-point difference implies a luminosity-weighted age difference
of only $\sim 1$ Gyr between the corresponding stellar populations, with
field galaxies being younger. The mass-weighted age difference could be
significantly smaller, if minor events of late star formation took place
preferentially in field galaxies.  We combine these results with the existing
evidence for the bulk of stars in cluster early-type galaxies having
formed at very high redshift, and conclude that the bulk of stars in
galactic spheroids had to form at high redshifts ($z\gsim 3$), no matter
whether such spheroids now reside in low or high density regions. The
cosmological implications of these findings are briefly discussed.
\end{abstract}


\section{Introduction}

Great progress has been made in recent years towards charting and
modeling galaxy formation and evolution. Yet, the origin of the
 galaxy morphologies, as illustrated by the Hubble classification, has
so far defied a generally accepted explanation. This is especially the
case for elliptical galaxies, with two quite different scenarios still
confronting each other. One scenario is motivated by hierarchical clustering
cosmologies, where ellipticals are modeled to form through a series of
merging events taking place over a major fraction of the cosmological
time (e.g. Baugh, Cole, \& Frenk 1996; Kauffmann 1996). 
The other scenario assumes instead the whole baryonic mass of the galaxy
 being already assembled at early times in gaseous form, and for this
reason it is sometimes qualified as {\it monolithic}. Early examples of
this latter scenario (Larson 1974; Arimoto \& Yoshii 1987) 
stemmed from the Milky Way collapse model of
Eggen, Lynden-Bell, \& Sandage (1962), and late realizations include
models by Bressan, Chiosi, \& Fagotto (1994) and Matteucci (1994).

Through the 1980's much of the debate focused on the age of
ellipticals as derived from the integrated spectrum of their stellar
populations. In general, advocates of the merger model favored an
intermediate age for the bulk of stars in ellipticals, but the matter
remained controversial (for opposite views see O'Connell 1986, and
Renzini 1986). A first breakthrough came from noting the very tight 
color-$\sigma$ relation followed by ellipticals in the Virgo and Coma
clusters (Bower, Lucey, \& Ellis 1992). This demostrated that at least
{\it cluster} ellipticals are made of very old stars, with the bulk
of them having formed at $z\gsim 2$. Evidence in support of
this conclusion has greatly expanded over the last few years. This
came from the tightness of the fundamental plane
relation for ellipticals in local clusters (Renzini \& Ciotti 1993),
from the tightness of the color-magnitude relation for ellipticals in
clusters up to $z\sim 1$ (e.g., Aragon-Salamanca et al. 1993; Stanford,
Eisenhardt, \& Dickinson 1998), and from the
modest shift with increasing redshift in the zero-point of the fundamental
plane, Mg$_2-\sigma$, and color-magnitude relations of cluster
ellipticals (e.g., Bender et al. 1997;
Dickinson 1995; Ellis et al. 1997; van Dokkum et al. 1998;
Pahre, Djorgovski, \& de Carvalho 1997; Stanford,
Eisenhardt, \& Dickinson 1998; Kodama et al. 1998). All these studies
agree in concluding that most stars in ellipticals formed at $z\gsim 3$. 

However, much of this evidence is restricted to cluster ellipticals.
In hierarchical models, clusters form out of the highest peaks in
the primordial density fluctuations, and cluster ellipticals completing
most of their star formation at high redshifts could be accommodated in
the model (e.g. Kauffmann 1996; Kauffmann \& Charlot 1998). However, in lower
density, {\it field} environments, both star formation and merging are 
appreciably delayed to later times (Kauffmann 1996), which offers the
opportunity for an observational test of the hierarchical merger
model.

The notion of field ellipticals being a less homogeneous sample compared
to their cluster counterparts has been widely entertained,
though the
direct evidence has been only rarely discussed.  Visvanathan \& Sandage
(1977) found cluster and field ellipticals to follow the same
color-magnitude relation, but Larson, Tinsley, \& Caldwell (1980) --
using the same database -- concluded that the scatter about the mean
relation is larger in the field than in clusters (see also Burstein 1977).
More recently, a larger scatter in field versus
cluster ellipticals was also found for the fundamental plane (FP) 
relations by de Carvalho \& Djorgovski (1992). However,
at least part of the larger scatter for the field ellipticals can be a
mere manifestation of the distances being more uncertain, which will also
affect the FP relations. Moreover, the database analyzed
by de Carvalho \& Djorgovki includes only $\sim 60$ cluster galaxies
and about the same number of field galaxies. One can also suspect most
of the effect being due to reddening errors (Pahre 1998).

The lack of conclusive evidence for or against systematic differences
between clusters and field  ellipticals has prompted  us to take
advantage of the large database assembled for the ENEAR project (da
Costa et al. 1998). The aim of the project is to create a homogeneous
database for a well-defined magnitude-limited  sample of early-type
galaxies in order to reconstruct the peculiar velocity and  mass density
fields out to a distance of $cz=7000$ km s$^{-1}$. Besides redshift, 
the measured
quantities include the central velocity  dispersion \so, the magnesium
index \mg2, and the photometric parameters $D_n$, $R_e$ and $\mu_e$. The
analysis of other absorption lines (H$\beta$, Fe, NaD) is currently
underway. Since both \so \ and \mg2 are distance and reddening
independent quantities, the comparison of the \mg2-\so \ relations for
cluster and field ellipticals offers the best available way of
establishing whether intrinsic differences exist between the two
populations. 

The paper is organized as follows. In Section 2 the samples of cluster
and field galaxies are defined,
and the corresponding \mg2-\so \ relations are presented and analyzed.
In Section 3 the results are interpreted and used to gather clues on
the formation of early-type galaxies.

\section{THE \mg2--\so \ RELATION IN CLUSTERS AND IN THE FIELD}

The ENEAR database includes over 2000 early-type galaxies with 
$m_{B(0)} \leq 14.5$
and $cz \leq 7000$ km/s, and galaxies in well-known nearby clusters 
which are  used to derive distance relations (for a full description
of the database see da Costa et al. 1998a).  While part of the ENEAR
database consists of data assembled from the literature, 
in the present analysis we restrict ourselves to the sample of 931
galaxies so far observed specifically for the ENEAR project 
(Bernardi et al. 1998a; Wegner et al.
1998). Among them, 232 galaxies have $T= -5$ (E Type), 189 have $T= -3$
(E-S0 Type), and the remaining 510 have $T= -2$ (S0 Type)  according to the
morphological classification of Lauberts \& Valentijn (1989). The vast
majority of these galaxies have disk to bulge ratios $D/B < 1$
(Alonso et al. 1998; Bernardi et al. 1998a). 
The spectra come
from a variety of telescopes (ESO 1.52m, MDM 2.4m and 1.3m and CASLEO 
2.15m) and instrumental setups.  
Special care was taken to cross calibrate against each other the \mg2
and \so\ measurements of spectra obtained with
different setups  (see Bernardi et al. 1998a for
details). After bringing the various measurements to a uniform system,
the calibration to the Lick system (Worthey et al. 1994) was enforced
using galaxies in common with the updated 7 Samurai (7S) sample
(Burstein 1998).  Typical internal errors (as well as differences with
other datasets, e.g. J\/orgensen et al. 1996) are $5\%$ to $13\%$ for
\so\ and $0.005$ to $0.011$ mag for \mg2.


We were especially careful in assigning galaxies to cluster,
group, and field environments.  We used the known clusters to obtain a combined
$D_{n}-$\so \ relation (Bernardi et al. 1998b) which was used to estimate
galaxy distances. Clusters were defined as those
aggregates containing at least 20 galaxies  and groups as those
including at least 2 and less than 20 members in the group
catalogs of  Ramella et al. (1997, 1998), which correspond to
overdensities of $\delta \rho / \rho \geq 80$. 
These catalogs derived from complete redshift surveys (CfA2, Geller \&
Huchra 1989; SSRS2, da Costa et al. 1998b).\\
Assignment to a cluster or group was then made for our early-type
galaxies fulfilling the following criteria: $d_{\rm i} \le 1.5 R_p$ and
$c|z_{\rm i}-z_{\rm cl}| \le 1.5 \sigma_{\rm cl}$, where $d_{\rm i} $
and $cz_{\rm i}$ are the distance from the cluster center and the radial
velocity of the galaxy, respectively, $R_{\rm p}$ is the pair radius
(Ramella et al. 1989), and $cz_{\rm cl}$ and $\sigma_{\rm cl}$ are
respectively the radial velocity and velocity dispersion of the cluster. When
applying these criteria a few galaxies originally assigned to the
clusters (and used to derive the $D_{n}-$\so \ relation) dropped out
of the sample, while a few new ones were included. The $D_{n}-$\so \
relation was then 
adjusted iteratively until convergence was reached.  In this way, 151
and 128 galaxies have been finally assigned to clusters and groups,
respectively. All the remaining galaxies were assigned to the field (631
objects), after having excluded a few close
pairs and those in the outskirts of clusters 
with $d_{\rm i} \le 3 R_p$ and $c|z_{\rm i}-z_{\rm cl}| \le 3
\sigma_{\rm cl}$, whose assignment was ambiguous.

The resulting \mg2-\so\ relations are shown in Fig. 1, for the whole
sample, as well as separately for the field, group, and cluster samples.
Also shown are linear least squares fits to the data  (\mg2 = $a{\rm
\log}$\,\so $ + b$),  where $a$ is the slope and $b$ the zero-point.
 For each subsample the slope obtained for the whole sample
was retained, and only the zero-point was derived. As it is evident from
the figure, field, group, and cluster ellipticals all follow basically
the same relation. The zero-point offset between cluster and field
galaxies is $0.007\pm 0.002$ mag, with field galaxies having lower
values of \mg2, a statistically significant, yet very small difference.
This is in excellent agreement with the offset of $0.009\pm 0.002$ mag,
obtained by J\/orgensen (1997) using 100 field and 143 cluster galaxies
from the {\it old} 7S sample (Faber et al. 1989). Our own
redetermination using the revised 7S sample (Burstein 1998) yields a
marginally lower value, i.e., 0.005$\pm 0.002$ mag. Fig. 2 shows a
histogram of the residuals for the ENEAR sample. The rms of the field
sample is 0.032 mag, virtually identical to that of the cluster sample
(0.030 mag). This is appreciably larger than our estimated internal
errors, indicating that most of the scatter is indeed intrinsic (cf.
Colless et al. 1998).

Subsamples of the cluster and field galaxies have been analyzed in
search of possible correlations. No significant correlations of the
residuals were found with morphology nor $D/B$ ratios.
In practice we recover here the result that ellipticals and spiral
bulges are alike (Jablonka, Martin, \& Arimoto 1996). Marginally
significant differences are instead found when dividing  about the
median each of the samples into high and low velocity dispersion
subsamples (at log$\,$\so=2.15), and high and low luminosity subsamples
(at $M_{\rm B_0}=-18.5$). (The subsamples are highly correlated given
the Faber-Jackson relation.) When keeping the slope constant, the zero
point difference between the high velocity/high luminosity cluster and
field subsamples is $0.005\pm 0.004$ mag. The difference between the low
luminosity/low velocity subsamples is instead $0.011\pm 0.006$ mag. If
anything, it appears that bright/massive galaxies form a more
homogeneous population, with a smaller difference in their \mg2-\so \
relation between cluster and field objects, compared to subsamples of
intrinsically smaller galaxies.  Finally, it is worth noting that no
correlation seems to exist between the zero-point of the \mg2--\so\
relation for cluster ellipticals in the EFAR sample and cluster richness
as measured  by cluster X-ray luminosity, temperature of the ICM, or
$\sigma_{\rm cl}$ (Colless et al. 1998). The present study extends this
(lack of) trend to the lowest density regions inhabited by early-type
galaxies.

\section{DISCUSSION AND CONCLUSIONS}

As is well known, the \mg2 index of a stellar population depends on both
age and metallicity. When dealing with real galaxies, it will also
depend on the detailed {\it distribution} of stellar ages and
metallicities within a given galaxy (Greggio 1997). Here we let this
complication aside (though it may help explaining the intrinsic scatter
of the \mg2-\so \ relation), and use simple stellar population models to
 set constraints on an indicative {\it age difference} between cluster
and field ellipticals. As widely appreciated, population models are
still affected by several limitations, which makes unwise to use them 
to determine the absolute age of galaxies. However, here we use 
the models only in a {\it differential} fashion, hence our conclusions
should be less prone to systematic errors that may affect such models.
For solar composition and an age in excess of 10
Gyr, the time derivative $(\partial$\mg2 $/\partial t)$ is 0.0060,
0.0034, and 0.0077 mag/Gyr, in the models of Buzzoni, Gariboldi, \&
Mantegazza (1992), Worthey (1994), and Weiss, Peletier, \& Matteucci 
(1995), respectively. A straight average gives $(\partial$\mg2
$/\partial t)\simeq 0.0057$ mag/Gyr, or $\Delta t({\rm Gyr})\simeq
175\,\Delta$\mg2. Therefore, the zero-point offset between cluster and
field galaxies suggests an average age difference between the two
samples of  $\sim 1.2\pm 0.35$ Gyr.  This roughly corresponds to a
luminosity-weighted age, while the actual, mass-weighted age difference
can be substantially smaller. To produce the observed offset it is
indeed sufficient  that some galaxies have undergone a minor star
formation event a few Gyr ago, and that this has taken place
preferentially among field galaxies (this effect may have been already
detected among HDF ellipticals, see Abraham et al. 1998). Therefore,
this $\sim 1$ Gyr age difference should be regarded as an upper limit to
the intrinsic, mass averaged age of stars in field and cluster
ellipticals. Of course, given the age/metallicity degeneracy affecting
spectroscopic indices such as \mg2, one can claim that the age
difference may be larger than the above limit, but is almost precisely
compensated by field galaxies being more metal rich at any given value
of \so.  We find this alternative interpretation very contrived, hence
unattractive.

We are now in a position to compare with theoretical simulations. In
the hierarchical merger model of Kauffmann (1996) the
luminosity-weighted age of stars in bright ellipticals that
reside in low-density environments is about 4 Gyr less than that of
cluster galaxies of similar luminosity. This would correspond to a difference 
$\Delta$\mg2$\simeq 0.023$ mag, which our data exclude at the
$4.6\sigma$ level. Indeed, in the hierarchical merger model, 
the brightest field
ellipticals form last (as expected) while smaller ones are instead
more coeval to cluster galaxies.  The evidence presented in
Section 2 suggests the opposite: brighter field galaxies appear to be
more similar to their cluster counterparts than the fainter ones.
We should warn that the specific model with which we are comparing refers to
a standard CDM model, i.e. $\Omega =1$. Hierarchical models for low $\Omega$
(and even more so $\Lambda$ models) should produce more homogeneous
populations of ellipticals and spheroids.
It remains to be seen whether such models can produce cluster and field
galaxies following the same \mg2-\so \ relations.

The present results do not necessarily invalidate the hierarchical merging
paradigm, but tend to push the action back to an earlier cosmological
epoch, favoring a scenario in which merging takes place at high
redshifts, among still mostly gaseous components in which the
merging itself promotes widespread starburst activity. The natural 
observational counterparts of these events is represented by the
Lyman-break galaxies at $z\gsim 3$ (Steidel et al. 1996), where star
formation rates can reach values as high as $\sim 1000\; \msun\yr-1$ 
(Dickinson 1998).

Combining the evidence mentioned in Section 1 of this paper, with the
 close similarity of cluster and field early-type galaxies documented
here, one can conclude that the bulk of stellar populations in galactic
spheroids formed at high redshift ($z\gsim 3$), no matter whether such
spheroids
now reside in high or low density regions. Additional direct evidence
supporting this conclusion also come from stellar color-magnitude
diagrams of globular clusters and fields in the bulge of our own
 Galaxy, that indicate a uniform old age for the Galactic spheroid
(Ortolani et al. 1995). With spheroids containing at least 30\% of all
stars in the local universe (Schechter \& Dressler 1987; 
Persic \& Salucci  1992) or even more
(Fukujita, Hogan, \& Peebles 1998), one can conclude that at least
 30\% of all stars and metals have formed at $z\gsim 3$ (Renzini 1998;
see also Dressler \& Gunn 1990).
This is several times more than suggested by a conservative
 interpretation of the early attempt at tracing the cosmic history of
 star formation, either empirically (Madau et al. 1996) or from 
theoretical simulations (e.g. Baugh et al. 1996). 
Yet, it is is more in line with
recent direct estimates from the spectroscopy of Lyman-break galaxies
 (Steidel et al. 1998).

We are indebted to Dave Burstein for having provided us with the
updated 7S sample and to Roberto Saglia for useful discussions and advice.
We are also grateful to the anonymous referee for constructive comments.
MB wishes to thank Wolfram Freudling for advice in the data reduction.
MVA acknowledges partial funding grants from CONICOR, CONICET and Fundaci\'on Antorchas (Argentina).
PSP acknowledges funding from CNPq grant 301373/86-8 and from
the Centro Latino-Americano de F\'\i sica.
GW was partially supported by NSF through Grant AST93-47714, and by 
the Alexander von Humboldt-Stiftung during 
stays at the Ruhr-Universit\"at (Bochum) and at  ESO (Garching).
CNAW acknowledges partial support
from CNPq grants 301364/86-9, 453488/96-0 and from the ESO Visitor
program.

\clearpage

\newpage
\begin{figure}
\plotone{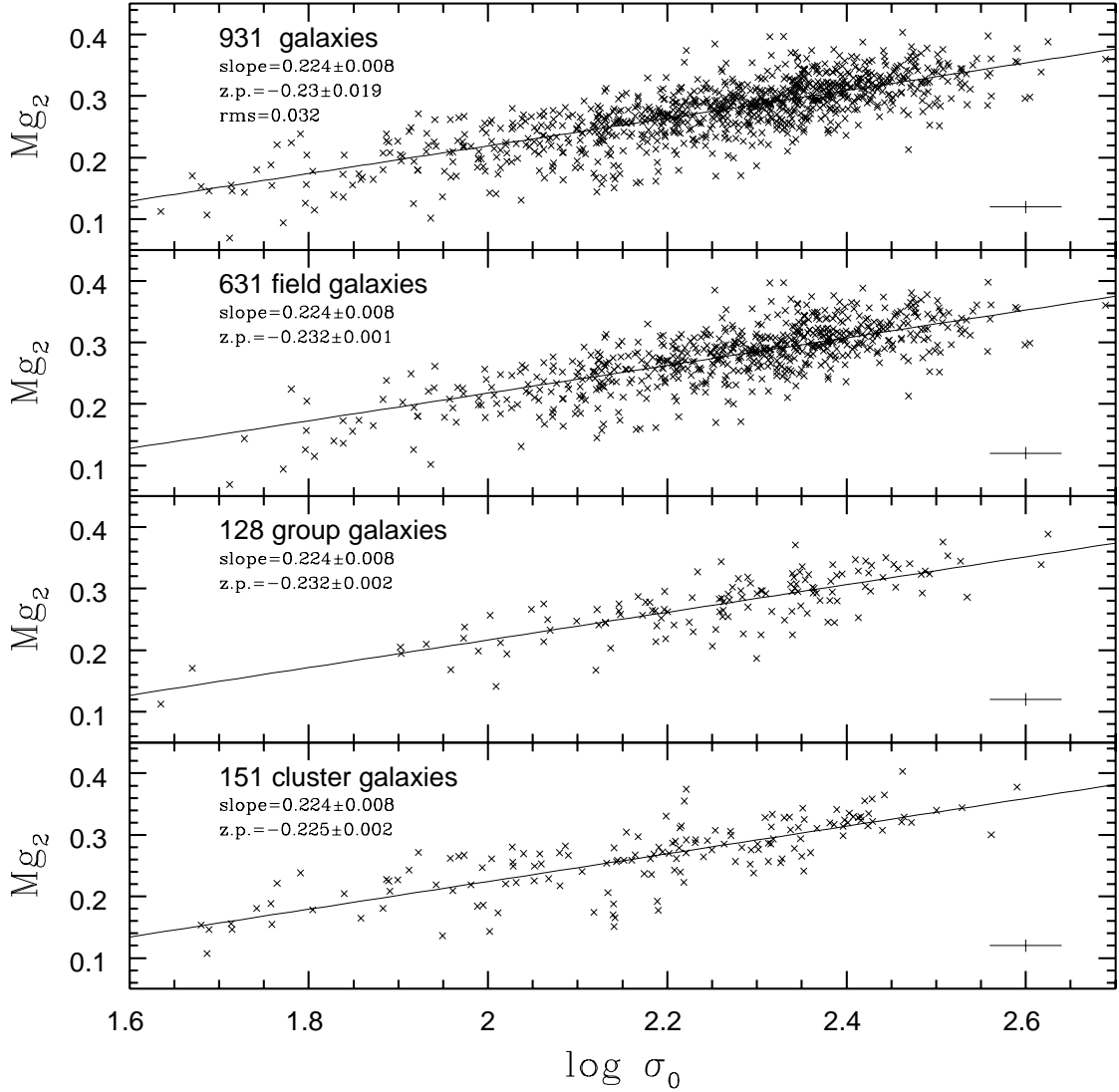}
\figurenum{1}
\caption{The \mg2--\so\ relation for the total sample of early-type
galaxies (upper panel), as well as for the field, group and cluster
subsamples (lower panels). The corresponding number of objects,
the slope, and the zero-point (z.p.) are shown in the upper
left corner of each panel. The least squares fits to the \mg2--\so\ relation
are also shown. For the three subsamples the slope as
derived for the total sample was retained, and only the zero-point was
determined. The error bars are shown in the lower right corner.}
\label{L2}
\end{figure}

\newpage
\begin{figure}
\plotone{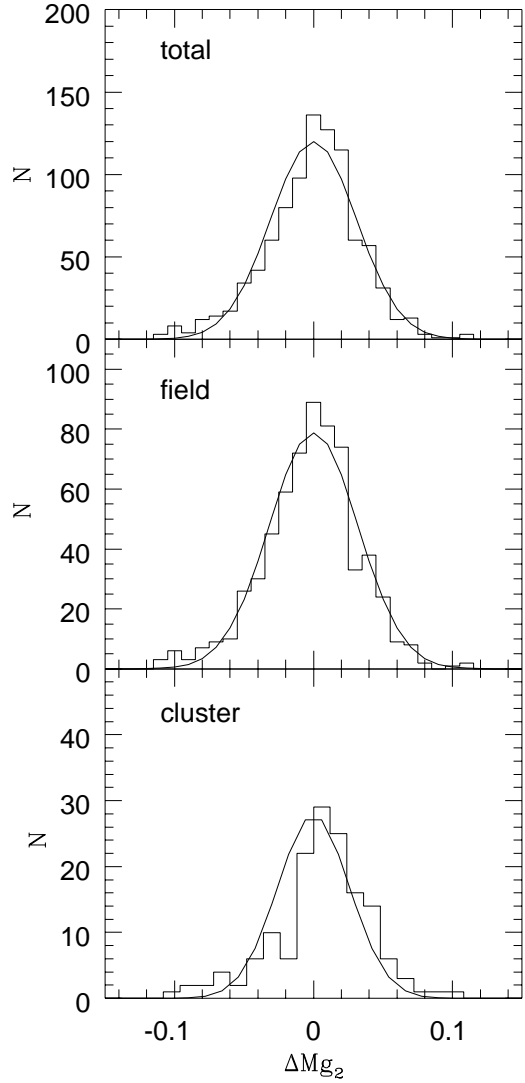}
\figurenum{2}
\caption{The distribution of the \mg2 residuals relative to the least 
squares fit obtained for the total sample in Fig. 1 are shown for the total, 
field and cluster data sets. 
The Gaussian bestfitting the residuals for the total sample is
overplotted in each panel.} 
\label{n4}
\end{figure}

\end{document}